# INTERCONNECTION OF COMMUNITIES OF PRACTICE: A WEB PLATFORM FOR KNOWLEDGE MANAGEMENT

Élise Garrot-Lavoué

*Université de Lyon, INSA Lyon, Laboratoire LIESP, 21 avenue Jean Capelle, F-69621 Villeurbanne, France*
*Elise.Garrot@insa-lyon.fr*

Abstract: Our works aim at developing a Web platform to connect various Communities of Practice (CoPs) and to capitalise on all their knowledge. This platform addresses CoPs interested in a same general activity, for example tutoring. For that purpose, we propose a general model of Interconnection of Communities of Practice (ICP), based on the concept of Constellation of Practice (CCP) developed by Wenger (1998). The model of ICP was implemented and has been used to develop the TE-Cap 2 platform which has, as its field of application, educational tutoring activities. In particular, we propose an indexation and search tool for the ICP knowledge base. The TE-Cap 2 platform has been used in real conditions. We present the main results of this descriptive investigation to validate this work.

## 1 INTRODUCTION

People belong to Communities of Practice (CoPs) at the local level of their company or institution. The emergence of such communities occurs when people have informal discussions, help each other to solve problems and use this to develop their competencies and expertise. CoPs centred on a same activity can have similar practices without being necessarily aware of it, mainly due to the fact that they do not belong to the same company or institution. As a result, every local CoP develops its own practices, each one reinventing what is certainly being replicated somewhere else. Our work is illustrated throughout the article by the example of tutoring, which we define as the educational monitoring of learners during courses. CoPs of tutors from different educational institutions prepare their own pedagogical contents for their students, and there is currently no possibility of reusing and sharing them (Garrot *et al.*, 2009). The result of this is that tutors lack help in their day-to-day practice, professional identity and practice sharing.

The problem which is challenging us is the creation of relation between local CoPs of actors practicing a same activity so that they exchange their knowledge and produce more knowledge than separate communities. We aim at developing a Web platform to capitalise on all produced knowledge by contextualising it, so as to make it accessible and reusable by all members in their working contexts.

Our work is based on the concept of Constellation of Communities of practice (or CCP) developed by Wenger (1998). In this article, we first present the main characteristics of this concept on which we base our research. We then situate our works by studying existing knowledge management systems and social networking services. In the third section, we propose a model of Interconnection of Communities of Practice (ICP), as an extension of the concept of CCP. This model approaches the actors' activity according to the point of view of interconnected practices and considers CoPs' members to act as the nodes between CoPs to support knowledge dissemination. In the fourth section, we present the implementation of the model of ICP by the development of the TE-Cap 2 platform, meant for CoPs of educational tutors from different institutions, countries and disciplines who would tutoring. We finally validate our works by presenting the main results of a descriptive investigation.

## 2 CONSTELLATION OF COMMUNITIES OF PRACTICE

Explaining that some organisations are too wide to be considered as CoPs, Wenger (1998) sets out his vision of these organisations as "Constellation of Communities of practice" (or "CCP").

We first define "Communities of Practice" (or "CoPs"). Communities of Practice gather people together in an informal way (Lave and Wenger, 1991) because of the fact that they have common practices, interests and purposes (i.e. to share ideas and experiences, build common tools, and develop relations between peers) (Wenger, 1998; Koh and Kim, 2004). Their members exchange information, help each other to develop their skills and expertise and solve problems in an innovative way (Pan and Leidner, 2003; Snyder *et al.*, 2004). They develop a community identity around shared knowledge, common approaches and established practices and create a shared directory of common resources.

We identify three main aspects of the concept of CCP, on which we base our works so as to develop a platform to support several Communities of Practice (CoP), summarised by Figure 1:

- **To favour interactions among CoPs**. Brown and Duguid (1991) brought the notion of "communities-of-communities" to develop the innovation within organisations, considering that the productions of separate communities can be increased by exchanges among these communities. The concept of Constellation of Communities of Practice (Wenger, 1998) resumes this idea by directing it on practices. The advantage to define several communities around shared practices is to create more knowledge and to develop more interactions than in a global community (Pan and Leidner, 2003). An involvement of this vision is to think about interactions among practices, rather than to favour information flows.
- **To consider the boundaries of CoPs as places of creation of knowledge**. The relations between communities can be supported by boundary objects (Star and Griesemer, 1989) and by brokering. Boundary objects are products of reification and they constitute the directory of resources shared by all the communities. Interactions between communities relate to this knowledge. "Brokers" belong to multiple communities and have a role of knowledge import-export between these communities. According to Ziovas and Grigoriadou (2007), the combination of brokering as a product of participation and the boundary objects as a product of reification is an effective way to create relations between CoPs. The meetings on the boundaries of CoPs arouse interactions between the members, what makes boundaries the places of creation of knowledge;
- **To establish a balance in the duality local/global**. A person belongs to and involves in one or several CoPs, each bound to its local practices. But the concept of constellation approaches the CoPs in a global point of view, as a set of practices negotiated with only one shared resources repository. Every member, as broker, operates the dissemination of knowledge from a level of practice to another one. That is why it is necessary to supply all CoPs with multiple means of communication between practices which feed the shared directory (Wenger, 1998).

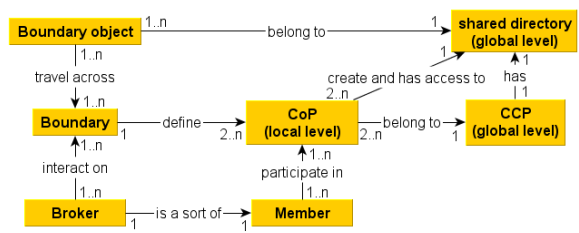

Figure 1: Modelling of the concept of Constellation of Communities of Practice (CCP).

## 3 KNOWLEDGE MANAGEMENT SYSTEMS AND SOCIAL NETWORKING SERVICES

In this section, we situate our works with regard to KM systems and social networking services so as to show that we cannot use existing complete solutions.

A KM system has to support the KM process following three stages (Von Krogh, 1999): capturing knowledge, sharing and transferring knowledge, generating new knowledge. The KM platform of a company is aimed at its organisational entities, what implies that:

- These systems are not designed to CoPs which do not correspond to traditional organisational entities;
- The proposed computer tools are the only means for the employees to communicate remotely; they thus have to use them if they want to exchange their practices;

- The employees meet during meetings within their organisational entities, so weave relations except the platform.
- The employees belong to organisational entities for which they already have a *feeling* of membership.

Since our works concern actors who do not necessary belong to the same institution or the same company, we cannot use an existing KM platform. The most important difficulty to overcome is to arouse interactions between persons except any frame imposed by an organisation. For that purpose, it is necessary to bring them to become aware that they have shared practices and to provide the available means to get in touch with people from different CoPs.

Some Web 2.0 applications as Facebook or MySpace are social networking services which "connect you with the people around you". They are very good examples of services which aim at connecting people who have common interests. Some social networking services are for more professional vocation, such as LinkedIn and Viadeo. But these sites are used for socialisation and to meet people. A consequence is that the tools offered to classify and to search for knowledge are not adapted to CoPs. Indeed, they often rest on collective categorisation in the form of tag clouds (O'Reilly, 2005) (folksonomies) or on full text search. But this system of 'tagging' lacks structuring (Guy and Tonkin, 2006). Within the framework of a CoP, we consider it is necessary to bring a knowledge organization to help users to index and search for knowledge. Tags systems work well for communities of interest where the users want to navigate within the application without precise intention. But these systems are not really adapted to CoPs where the users search for resources bound to working experiences. Users must be able to find a testimony, a discussion, an 'expert' or other resources (document, Web link…) very quickly, so that they can use it in their practice.

To sum up, we can use neither complete KM solutions nor existing social networking services but we can use existing components. We adopt one of the Web 2.0 principles: "innovation in assembly" (O'Reilly, 2005). When there are a lot of basic components, it is possible to create value by assembling them in a new way. We chose to develop a platform partially composed of existing Web 2.0 tools (Wenger *et al.*, 2005), available as well for KM systems as for social networking services, to capitalise knowledge and get in touch with people. Other part of the system consists of a knowledge indexation and search tool specifically developed to answer specific needs of CoPs, based on the model on Interconnection of Communities of Practice depicted in next section.

## 4 MODEL OF INTERCONNECTION OF COMMUNITIES OF PRACTICE

The concept of CCP is based on the assumption that considering a global community as a set of interconnected CoPs increase member participation and creation of knowledge. Furthermore, this vision of an organisation takes into account as much the local level of every CoP as the global level formed by all the CoPs. We adopt this approach to develop a model of Interconnection of CoPs (ICP) which proposes to approach a general activity according to multiple points of view depending on actors' practices. The development of the Web platform Te-Cap 2, depicted in section 5, is based on this model.

### 4.1 General Model of ICP

In the case of informal professions, such as tutoring, it is difficult to define exactly the field of practice of the actors. Actors' activities can be seen as a set of different practices which are similar in some points. For example, tutors' roles can be different as their interventions could be punctual or long-lasting; the learning session could be computer mediated or not and the learners' activity could be individual or collective. But some roles are shared by some of these contexts. We propose that this group of actors should be seen not as an endogenous entity defined by a field of practice, but rather as *a set of CoPs supported by a Web platform where individual members acting as nodes of interconnected practices are the connection points* (see Figure 2). We suggest developing this concept that we have named Interconnection of Communities of Practice (ICP). This model aims at making existing local CoPs of actors (e.g. within an educational institution), who are engaging in the same general activity (i.e. tutoring), to get connected. This model also proposes active support for the dissemination of knowledge from CoP to CoP.

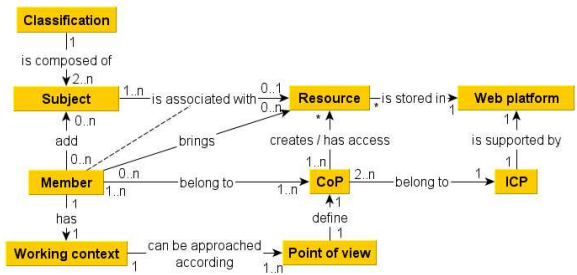

Figure 2: General model of Interconnection of Communities of Practice.

At an individual level, an actor's activity can be approached according to multiple points of view depending on the working context. In the ICP model, a CoP corresponds to the elementary level of actors' practice. The CoPs to which they belong are defined by their working context. At a general level, an ICP is composed of all the elementary CoPs defined by all the actors who participate in the Web platform. We could see it as a single community of actors practicing a same activity, brought together on the same platform; a group which can be approached from multiple points of view and accessed through multiple entry points.

For example (see Figure 3), Tutor 1, working in the industrial engineering department of the University A in France who is monitoring a collective project about maintenance can belong to five different CoPs: tutors who monitor collective activities, tutors who are interested in maintenance, tutors who monitor educational projects, tutors of the industrial engineering department and tutors of the University A. Tutor 2 from another educational institution, for example University B in Canada, can belong to several CoPs, some of which Tutor 1 may also belong to.

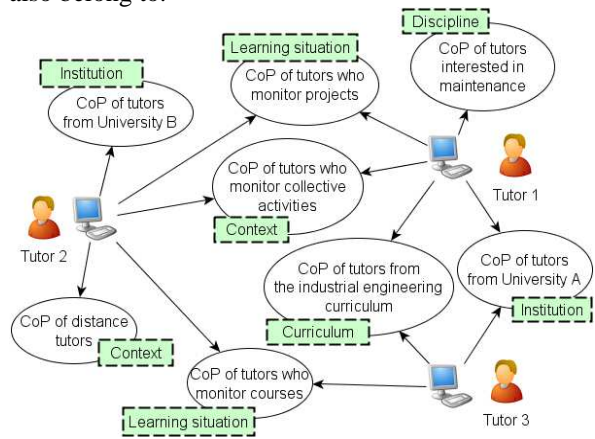

Figure 3: Tutors as nodes of Interconnection of CoPs.

These two tutors, from different countries, will be put in touch since their working context can be approached according to several similar points of view, which imply that they belong to same CoPs. Tutor 3 will be put in touch with both tutors because he belongs to the same educational institution and the same department as Tutor 1 and because he monitors the same type of activity as Tutor 2. So this example illustrates the fact that it is the tutors who are the nodes of Interconnection of CoPs. In this example, tutors' activity can be approached from several points of view: the context of the activity (collective, distance), the learning situation (project based learning, courses), the discipline (maintenance), the curriculum (industrial engineering) and the educational institution (universities). These points of view are categories of CoP and we propose in section 4.3 an approach to define a model of actors' practices, which implies determining all the categories of CoPs and which CoPs correspond to a given activity.

## 4.2 The reasons for using ICP instead of CCP

We based the model of ICP on the model of CCP since they suggest both considering wide organisations as a set of communities of practice which have common characteristics (Wenger 1998):
- They share members: the ICP members belong to several CoPs, each corresponding to a point of view of their working context;
- They share artefacts: the ICP members participate on the same Web platform;
- They have access to the same resources: the ICP members have access to the shared directory of resources stored in the platform database.

However, an organisation defined as an Interconnection of CoPs (supported by a Web platform and composed of individual members who act as nodes of interconnected practices) does not form a Constellation of CoPs as defined by Wenger:
- Contrary to a CCP, the CoPs of an ICP do not share historic roots on which the mutual engagement of the members could base itself. The ICP members do not know apart the platform on which they join. This difference is fundamental because it raises the difficulty bringing persons who do not know each other to interact, what requires supporting a high level of sociability on the platform.
- In a CCP, the CoPs have interconnected projects which connect them whereas an ICP

consist of actors practicing a same general activity who want to exchange on their practices with others, the community emerging by "propagation". So that members are interested in the practices of the others, it is important to bring them to be aware that they have rather close practices which they can share.
- Contrary to a CCP, the ICP members do not belong necessarily to the same institution. Since we aim at supporting exchanges as well in members' local working context as at the general level of the activity, it is necessary that there are actors of various institutions.
- The CoPs of a CCP are in close proximity to each other, in particular geographically, whereas an ICP is constituted of persons who meet themselves on a Web platform and can thus be from countries of the whole world. This model does not thus include geographical proximity.

So, we propose a new model of ICP to represent a close but different type of organisation which could be seen as:
- An **extension** of the model of CCP in the sense that the conditions are less restricting. We showed that only three conditions on seven put by Wenger (1998) are necessary to validate the existence of an ICP.
- A **transposition** of the model of CCP in the sense that it concerns persons gathered by a Web platform and not by a given institution or company.

### 4.3 Management and Dissemination of the ICP Knowledge

The ICP resources are stored in a database according to a hierarchical classification composed of subjects based on a model of actors' practices. In the case of tutoring, resources correspond to explicit knowledge (documents and Web links) and tacit knowledge shared among members (e.g. exchanges of experience, stories, and discussions). We built a model of tutors' practices which defines at most four levels. The first level corresponds to the main factors which differentiate actors' practices (e.g. educational institution, curriculum, discipline, activity) and are the main categories of CoP. Each category is divided into subcategories and so on. The terminal nodes correspond to CoPs. This taxonomy of tutoring has been developed by an iterative process, based on interviews with six tutors (first development cycle) and on results of an experiment of a first prototype (second development cycle). The classification cannot be exhaustive because it is only a base which will evolve through modifications and additions made by the ICP members themselves.

When creating a resource (message, document, Web link), the author decide that it belongs to one or several CoPs by associating the name of the CoP (subject in the lowest level of the classification) with the resource. When they find a resource (result of a search), members can also associate new subjects with this resource so as to spread it to new CoPs. They can either associate the name of a CoP to spread the resource to only a single CoP, or associate it to the name of a category of CoPs (subjects at higher levels in the classification) to spread the resource to all child CoPs. Indeed, Child CoPs (hierarchically lower level CoPs) inherit all the resources of a category of CoPs. So, ICP members' participation not only consists of creating new resources but also of creating links between these resources according to their relevance to the CoPs. This relevance is estimated by members themselves who consider a resource to be useful or interesting for a CoP. The supply of a resource to a CoP can lead to a debate on this resource and possibly to the creation of new resources for this CoP. Events reported in a precise context can lead to experience sharing (solutions, cases, scenarios), being used as a base to generate rules or recommendations which become global knowledge within the ICP.

## 5 THE TE-CAP 2 PLATFORM

We have developed the TE-Cap 2 (Tutoring Experience Capitalisation) platform according to a co-adaptive approach based on an iterative process including three development cycles. Each cycle rests on the development of a prototype, on its evaluation by the users by means of interviews or experiments and on the analysis users' activity (Garrot *et al.*, 2009). This approach aimed at making users' needs emerge, at leading users to explicit these needs. The platform specifications evolved according to these emerging needs. We were particularly interested in developing a knowledge indexation and search tool for an ICP. We describe this tool in the following section.

### 5.1 User Profile Management

The knowledge indexation and search tool is based on the user profiles used to personalise subjects proposed to them. Users define their profile by filling several fields corresponding to categories of CoPs of the hierarchical classification. Values given to fields define CoPs and imply tutors' membership of these CoPs. The profile is composed of three main characteristics: identity profile, working context and secondary interests. The working context is about all the CoPs directly bound to actors' working context. The secondary interests are about all the CoPs which are not directly bound to their working context but which could interest them (give access to other resources able to interest them and to profiles of other people who share similar practices or experiences).

As a tool provided for the use of members of a CoP in their daily practice, this one offers them fast access to the relevant resources for them by two means (see Figure 4):

- A link between the search interface and the profile allows users to only see the subjects from the classification which concern users and which interest them according to their profile. So users only have access to the resources of the CoPs to which they declare themselves to belong and can create resources only for these CoPs.
- Users have the possibility, according to their intention when connecting to the platform, to apply a filter to display on the classification interface only those subjects bound to their working context or to their secondary interests. In their daily practice, it is advisable to offer users at first only those subjects which concern their direct working context, this being the most efficient. If users do not find the information they look for in their direct working context, they must be able to extend the search to the other subjects of interest bound to their activity. In this manner they can find interesting 'unexpected' resources, which they can then bring into CoPs in which they have a central role.

## 5.2 Knowledge Indexation and Search Tool

The knowledge search and indexation tool, illustrated by Figure 4, rests on the classification built for the ICP. The main panel (at the centre of the screenshot), composed of three tabs, allows easy and fast navigation between the results of the search and the classification. The tab 'Search' gives the possibility of navigating within the classification and of selecting search subjects. These subjects are represented in the form of bubbles, to bring conviviality and attractiveness to the interface. Users can navigate in the classification by a 'double-click' on a bubble which explodes it into more bubbles representing the sub-subjects. When reaching the last level (corresponding to the CoPs), subjects are represented in the form of a combo box allowing a multiple selection. Users can return to a superior level thanks to the navigation path. The platform proposes the same interface to search for posted messages and for member profiles, by separating them by the way of two tabs. In this way users can, at every search, consult the profiles of found members and 'discover' people who have similar practices or who offer expertise.

The secondary panel (on the right of the screenshot) gives the possibility of storing the subjects chosen for the search (by a drag and drop from the main panel). The subjects in this column are always visible when users navigate in the tabs of the main panel and from one request to another. Once in the "search column" users can deselect or select a subject (so as to refine or to widen the search), delete a subject by sliding the bubble outside the column and move bubbles inside the column to choose a preferred order. This principle of category selection can be compared to carts on commercial Web sites. This original human computer interaction has been chosen to promote navigation within the classification and to simplify the selection of items.

The indexing of an initiating message (starting a discussion) is made according to the following principle: users classify the message according to its context (bound subjects) at the same time as they write it. This principle aims at leading them to reflect upon the experience they relate. To facilitate this action, an interface in the form of tabs ensures an easy navigation, at any time, between the writing and the indexing of a message. The selected subjects in the classification column are then associated with the message, meaning that this resource belongs to the CoPs or categories of CoPs. Every user can associate the discussion with new subjects so as to spread the resource from one CoP to another one and from one level to another. Regulation is carried out by the author of the initiating message who has the right to remove the subjects which they do not consider relevant for the discussion.

## 5.3 Classification Evolution

Users can make the resource classification evolve through their participation on the platform, so as to lead to a classification using a vocabulary which gradually moves closer to the actors' practices. For that purpose, the interface gives at any time the possibility of adding a new subject to the classification, be it when filling in a profile, when classifying a resource, when searching a resource or when consulting a resource. The subjects used are recorded which allows for example the deletion of those considered useless. Unused subjects are later deleted, meaning that they were not adapted to the actors' field of practice or not located at the right level of the classification. This evolution of subjects is necessary so that the classification made *a priori* becomes closer to the reality of actors' practices and can follow the evolution of actors' uses and practices. It is also an important point for ensuring a coherence of all the CoPs forming the ICP and for offering a common identity to all the members.

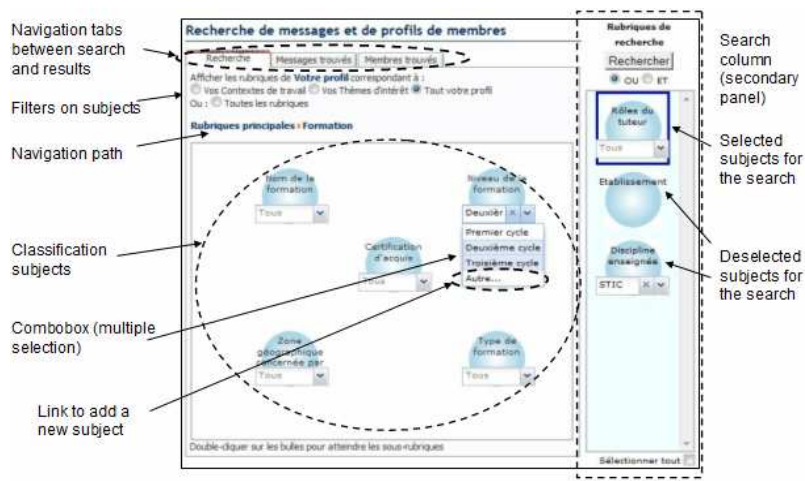

Figure 4: Knowledge search tool.

# 6 A DESCRIPTIVE INVESTIGATION

We conducted a descriptive investigation in real conditions, from 25 February 2008 to July 5 2008. Our role consisted of encouraging registered tutors to participate by sending regular newsletters. The Web address of TE-Cap 2 was disseminated to several communities of tutors (ATIEF, t@d, PALETTE) and to virtual campus (VCiel, FORSE, E-Miage, Téluq, Master UTICEF, did@cTIC, FLE). We also sent an email to the users of the first prototype TE-Cap. We wanted to develop the community around this existing core, hoping that they would encourage new users to participate. Discussion threads created during the first study were kept to be used as a base for new discussions.

To help in the understanding of the how the platform works, we posted online demonstration videos: one general one and three specific ones (how to do a search, to write a message and to fill in the profile). This study aimed at validating the TE-Cap 2 platform as a support for the interconnection of CoPs of tutors. We defined indicators to measure sociability, levels of knowledge creation and sharing and utility of the platform (Garrot, 2008). Results come from three types of data: use tracks, answers to a questionnaire and usability tests.

42 persons from nine francophone countries registered on TE-Cap 2. We present in this paper only the main results regarding the indexation and search tool. First of all, the answers to the questionnaire show that our aim to put local CoPs and online CoPs (general CoPs) into a relation answers an existing need. Indeed, tutors look for information as much at the local level of their course (eight answers to the questionnaire) as at a more general level such as tutors' roles (twelve answers), technical and educational tools and resources (twelve answers), learners (ten answers) or learning scenarios (eight answers). Although quite a few messages were written (fifteen) more (twenty-seven) users simply viewed discussions. This is explained by the fact that, according to the answers to the questionnaire, the users registered as much out of curiosity regarding a new tool as out of a desire to really participate. Furthermore, participation in a community will always be lower priority than teaching or tutoring. A positive result is the rather large number of subjects added to the classification (45), which implies a significant evolution in the classification and thus an appropriation by the users. Finally, usability tests carried out with three tutors according to a scenario, highlight the fact that the indexation and search interfaces of TE-Cap 2 are very easy to use and effective. But the use of these interfaces requires a learning stage, as is normal for an innovative interface which proposes new features. Furthermore, twenty-three users did not fill in or did not use their profile which, we must assume, means they not did not see the interest or did not take the time (it requires 5 to 10 minutes). The emphasised reason according to the questionnaire responses was that they did not understand the link between the profile and the proposed classification. It would be necessary to explain this link better so that they could see the interest. The help brought by the videos was either not sufficient or not adapted and an improvement could be the addition of a contextual help or a software companion.

Further results will be obtained only by a use by a large number of persons and over a longer time period. It is only in these conditions that the platform and the proposed tools can be expected to reveal their potential.

# 7 CONCLUSIONS

In this paper, we defined a general model of Interconnection of CoPs, based on the concept of Constellations of CoPs. This model aims at supporting knowledge sharing and dissemination for local CoPs interested in a same general activity, in our case tutoring. We validated the implementation of this model by the development of the TE-Cap 2 platform. This platform was designed to connect several CoPs centred on same general activity and to manage their knowledge. We conducted a descriptive investigation lasting several months with tutors from various disciplines and countries. The results of usability tests demonstrated the ease of use and the utility of the proposed tool, although not all the offered possibilities were taken up, as highlighted by use tracks.

The aim of this study was not to observe the emergence of an Interconnection of Communities of Practice because it was unachievable in only four months. So as to observe such emergence, we plan to conduct another type of study, across a long-term period and with the addition of a software companion to facilitate the understanding of the innovative interface. It would also be interesting to address other communities than that of tutors or teachers who often tend towards rather individualistic professional behaviour and who are not always used to sharing.

# REFERENCES

Brown, J.S. and Duguid, P., 1991. Organizational learning and communities of practice. *Organization Science*, 2(1), 40-57.

Garrot, E., George, S. and Prévôt, P., 2007. The Development of TE-Cap: an Assistance Environment for Online Tutors. *2nd European Conference on Technology Enhanced Learning (EC-TEL 2007)*, Lecture Notes in Computer Science, Springer, Crete, Greece, 481-486.

Garrot, E., George, S. and Prévôt, P., 2009. Supporting a Virtual Community of Tutors in Experience Capitalizing. *International Journal of Web Based Communities*, 5(3), 407-427.

Guy, M. and Tonkin, E., 2006. Folksonomies: Tidying up Tags? *D-Lib Magazine*, 12(1). Available at: http://www.dlib.org/dlib/january06/guy/01guy.html.

Koh, J. and Kim, Y., 2004. Knowledge sharing in virtual communities: an e-business perspective. *Expert Systems with Applications*, 26(2), 155–166.

Lave, J. and Wenger, E., 1991. *Situated Learning. Legitimate Peripheral Participation.* Cambridge, UK: Cambridge University Press.

O'Reilly, T., 2005. What Is Web 2.0: Design Patterns and Business Models for the Next Generation of Software. *O'Reilly Media*. Available at: http://www.oreillynet.com/pub/a/oreilly/tim/news/2005/09/30/what-is-web-20.html.

Pan, S. and Leidner, D., 2003. Bridging Communities of Practice with Information Technology in Pursuit of Global Knowledge Sharing. *Journal of Strategic Information Systems*, 12, 71-88.

Snyder, W.M., Wenger, E. and de Sousa, B.X., 2004. Communities of Practice in Government: Leveraging Knowledge for Performance. *The Public Manager*, 32(4), 17-21.

Star, S.L. and Griesemer, J.R., 1989. Institutional Ecology, `Translations' and Boundary Objects: Amateurs and Professionals in Berkeley's Museum of Vertebrate Zoology, 1907-39. *Social Studies of Science*, 19(3), 387-420.

Von Krogh, G., 1999. *Developing a knowledge-based theory of the firm*, St. Gallen, Switzerland: University of St. Gallen. Available at: http://www.dialogonleadership.org/docs/vonKrogh-1999.pdf.

Wenger, E., 1998. *Communities of practice: Learning, meaning, and identity*, Cambridge: Cambridge University Press.

Wenger, E., White, N., Smith, J.D., Rowe, K. and CEFRIO, 2005. Technology for Communities. *Guide to the implementation and leadership of intentional communities of practice. Work, learning and networked*. 71-94.

Ziovas, S. and Grigoriadou, M., 2007. Boundary Crossing and Knowledge Sharing in a Web-Based Community. *IADIS Web Based Communities Conference*. Salamanca, Spain, 248-256.